\documentstyle[11pt]{article}

\begin{document}
\begin{flushright}
CUMT-MATH-9503\par
NWU-IMP-95xx
\end{flushright}
\vspace{.5cm}
\centerline{\bf\Large{\bf{On the Classical $W_{4}^{(2)}$ Algebra}}}
\vspace{1in}
\centerline{\large L. \ Chao}
\smallskip
\smallskip
\centerline{ CCAST(World Laboratory),  P.O.Box 8730, Beijing 100080, China.}
\smallskip
\smallskip
\smallskip
\centerline{Institute of Modern Physics,  Northwest University,}
\centerline{Xian 710069, China}
\bigskip
\centerline{\bf\large and}
\bigskip

\centerline{\large Q.\ P. \ Liu}
\smallskip
\smallskip
\centerline{CCAST(World Laboratory), P.O.Box 8730, Beijing 100080, China.}
\smallskip
\smallskip
\smallskip
\centerline{Department of Mathematics, Beijing Graduate School,}
\centerline{China University of Mining and Technology, }
\centerline{ Beijing 100083, China\footnote{Mailing Address}}

\newcommand{\be}{\begin{equation}}
\newcommand{\ee}{\end{equation}}
\newcommand{\ba}{\begin{array}}
\newcommand{\ea}{\end{array}}
\renewcommand{\theequation}{\thesection.\arabic{equation}}
\def\w42{$W^{(2)}_4$}
\def\p{\partial}
\def\alf{\alpha}
\def\bi{\beta}
\def\es{\epsilon}
\def\la{\lambda}
\def\dl{\delta}
\vspace{0.5in}\begin{center}
\begin{minipage}{5in}
{\bf ABSTRACT}\hspace{.2in}We consider the classical \w42 algebra from
the integrable system viewpoint. The integrable evolution equations
associated with the \w42 algebra are constructed and the Miura maps
, consequently modifications, are presented. Modifying the Miura maps,
we give a free field realization the classical \w42 algebra. We also
 construct the Toda type integrable systems for it.\par
\end{minipage}

\end{center}
\vspace{.5in}
\vfill\eject
\par
\section{INTRODUCTION}
\setcounter{equation}{0}
Integrable systems of nonlinear diffenertial equations are studied
extensively during last three decades\cite{Dic}. These are the equations
which possess remarkable analytical, geometric and algebraic properties.
Evermore remarkable, this theory brings a number of different research
fields together and finds applications in several branches. The
interaction between integrable system theory and W algebra theory is
just one of the fascinating points, which attracts much attention recently
\cite{feher2}\cite{zam}\cite{ge}.
\par
$W_n$ algebra, which is higher spin generalization of Virasoro algebra,
is introduced by Zamolodchikov \cite{zam} recently. It plays an important
role in the theory of the two dimensional quantum gravity and matrix models.
$W_n$ algebra may be constructed via Hamiltonian reduction approach from
the WZW model\cite{feher2}. Novel W algebras exist which involves fields
with fractional spins. These are referred to $W_{n}^{(l)}$ algebras. The
first such example is the $W_{3}^{(2)}$ algebra of the Polyakov-Bershadsky
\cite{poly}, which consists of four fields: energy - momentum tensor,
two bosonic fields of spin $\frac{3}{2}$ and a spin 1 U(1) current.
\par
Gervais\cite{ge} is the first to notice the interrelation between the
KdV equation and the Virasoro algebra. Precisely, the second Poisson
bracket of the KdV equation is equivalent to the classical Virasoro
algebra. This result is generalized later on and the equivalence of
the classical $W_n$ algebra and the second Poisson bracket of the
Gelfand-Dickey hierarchy is discovered\cite{pm}\cite{bak1}. This
remarkable connection provides new insight into both theories. For
example, constructing new type of integrable systems may lead to new type
of W algebras, and vice versa. Another important point is that Miura
map, which plays a central role in the Soliton theory, often provides
free field realization for the corresponding W algebra. Noticing the
connection of the W algebras and integrable evolution equations, it
is not surprising that a correspondence between the W algebras and Toda
type system exists. We refer the papers\cite{feher2} for more details.
\par
The present paper is on the \w42 algebra. Generally speaking, $W_{n}^{(l)}$
algebras are given by different sl(2) embedding into sl(n). The form
of the \w42 algebra is inferred in \cite{Bais}, but its explicit form
is presented by Bakas and Depireux\cite{Bak} by means of Hamiltonian
reduction method. We notice that Mathieu and Depireux\cite{de} discussed
the $W_{n}^{(l)}$ algebra from the integrable system viewpoint, but
they failed to construct the integrable systems associated with the
classical \w42 algebra. We construct such systems in this paper.
Both nonlinear evolution equations and Toda type of systems will be
given explicitly for the \w42 algebra. Also, by working out Miura maps,
we obtain a free field realization for the classical \w42 algebra.
A by-product is that the free field realization for the W algebra
associated with matrix Schordinger operator.\par

The paper is arranged as follows. We recall the explicit form of the
classical \w42 algebra in the next section.  In section three, we
construct the hierarchy of nonlinear evolution equations for this
algebra. Section four is intended to construct its free field
realization. The integrable systems of Toda type are presented
in section 5. Final section contains some comments.\par

\section{THE CLASSICAL \w42 ALGEBRA}
\setcounter{equation}{0}
We recall the classical \w42 algebra in this section. It is presented
by Bakas and Depireux\cite{Bak} by means of Hamiltonian reduction
approach. It reads
\be
\begin{array}{ll}
\{T(x),T(y)\}=(\p^3+T\p+\p T)\delta(x-y),&\\[1mm]
\{T(x), v(y)\}= ({1\over2}\p^3-{3\over2}H\p^2+v\p +\p v+w\p p-
p\p w-H_{xx}-2H_x\p )\delta(x-y),&\\[1mm]
\{T(x), q(y)\}= ({1\over2}\p^2 w+q\p +\p q )\delta(x-y),~~~~~
\{T(x), r(y)\}= ({1\over2}\p^2 p+r\p +\p r )\delta(x-y),&\\[1mm]
\{T(x), w(y)\}= w\p \delta(x-y),~~~~~
\{T(x), p(y)\}= p\p \delta(x-y),~~~~~
\{T(x), H(y)\}= H\p \delta(x-y),&\\[1mm]

\{v(x),v(y)\}=({3\over4}\p^3-{3\over4}H\p H+v\p+\p v-
{3\over4}H_{xx}-{3\over2}H_x\p)\dl(x-y),&\\[1mm]

\{v(x),q(y)\}=({3\over4}\p^2 w +{3\over4}H\p w+\p q+Hq+
wv)\dl(x-y),&\\[1mm]

\{v(x),r(y)\}=(-p\p^2-{1\over4}\p^2 p+p\p H-pv-{1\over4}H\p p+
r\p -rH)\dl(x-y),&\\[1mm]

\{v(x),w(y)\}=-q\dl(x-y),~~~~~~~\{v(x),p(y)\}=(-p\p +r)\dl(x-y),&\\[1mm]

\{v(x),H(y)\}=\frac{1}{2}(-\p -H)\p \dl(x-y),~~~~~

\{q(x),q(y)\}=-{3\over4}w\p w\dl(x-y),&\\[1mm]

\{q(x),r(y)\}=(\p^3-\p^2 H-H\p^2 +\p v+u\p -H(u+v)+H\p H+
{1\over4}w\p p)\dl(x-y),&\\[1mm]

\{q(x),w(y)\}=0,~~~~~~~\{q(x),p(y)\}=((\p -H)\p -v+u)\dl(x-y),&\\[1mm]

\{q(x),H(y)\}=(\frac{1}{2}w\p -q) \dl(x-y),~~~~~~~

\{r(x),r(y)\}=-{3\over4}p\p p\dl(x-y),&\\[1mm]

\{r(x),w(y)\}=((\p -H)\p +v-u)\dl(x-y),~~~~~~~
\{r(x),p(y)\}=0,&\\[1mm]

\{r(x),H(y)\}=\frac{1}{2}(-p\p +r) \dl(x-y),~~~~~~~

\{w(x),w(y)\}=0,&\\[1mm]

\{w(x),p(y)\}=-2(\p-H)\dl(x-y),~~~~~~~
\{w(x),H(y)\}=-w \dl(x-y),
&\\[1mm]
\{p(x),p(y)\}=0,~~~~~~~\{p(x),H(y)\}=p\delta(x-y),
{}~~~~~~~\{H(x),H(y)\}=-\p\delta(x-y).&\\
\end{array}
\ee
We easily see that the algebra (2.1) is the one presented
in\cite{Bak} up to an invertible transformation.

\section{THE INTEGRABLE HIERARCHY OF EVOLUTION EQUATIONS}
\setcounter{equation}{0}
In order to derive an integrable hierarchy of nonlinear evolution
equations associated to the \w42 algebra, we specify the associated
spectral problem first. Our spectral problem is
\be
 \Phi_x= \left[\ba{cccc}  0&0&-\la&0\\ 0&0&0&-\la\\ u&q&H+h&0\\
r&v&p&-H+h\ea \right]\Phi \equiv U\Phi,
\ee
we adjoin (3.1) as usual with the time evolution of the wave
function $\Phi$: $\Phi_t=V\Phi$, then calculating the zero
curvature equation: $U_t-V_x+[U,V]=0$. By suitable adjustment,
 we find that the hierarchy related with (3.1) is represented as
\be
f_{t_{n}}=\{f,{\cal H}_{n+1}\}_0=\{f,{\cal H}_n\}_1,
\quad f=u, v, q, r, p, w, H, h.
\ee
where two Poisson brackets are defined by
\be
\begin{tabular}{ll}

$\{u(x),u(y)\}_0=-2\p \dl(x-y),$ &$\{u(x),q(y)\}_0=w \dl(x-y),$ \\[1mm]
$\{u(x),r(y)\}_0=-p\dl(x-y),$&$\{v(x),v(y)\}_0=-2\p \dl(x-y),$\\[1mm]
$\{v(x),q(y)\}_0=-w\dl(x-y),$&$\{v(x),r(y)\}_0=p\dl(x-y),$\\[1mm]
$\{q(x),r(y)\}_0=-2(\p+H)\dl(x-y),$& all other brackets vanish.\\
\end{tabular}
\ee
and
\be
\begin{array}{ll}
\{u(x),u(y)\}_1=(\p^3-(H+h)\p (H+h)+u\p+\p u+(H+h)_{xx}+
2(H+h)_x\p)\dl(x-y),&\\[1mm]
\{u(x),v(y)\}_1=(-w\p p-wr+qp)\dl(x-y),&\\[1mm]
\{u(x),q(y)\}_1=(-w\p^2-w\p (H+h)-wu+q\p +q(H+h))\dl(x-y),&\\[1mm]
\{u(x),r(y)\}_1=((\p -H-h)(\p p +r)+up)\dl(x-y),&\\[1mm]
\{u(x),w(y)\}_1=(-w\p +q)\dl(x-y),~~~~~~~
\{u(x),p(y)\}_1=-r\dl(x-y),&\\[1mm]
\{u(x),H(y)\}_1=\frac{1}{2}(\p -H-h)\p \dl(x-y),~~~~~
\{u(x),h(y)\}_1=\frac{1}{2}(\p -H-h)\p \dl(x-y),&\\[1mm]

\{v(x),v(y)\}_1=(\p^3-(H-h)\p (H-h)+v\p +\p v-(H-h)_{xx}
-2(H-h)_x\p)\dl(x-y),&\\[1mm]
\{v(x),q(y)\}_1=((\p +H-h)(\p w +q) +wv)\dl(x-y),&\\[1mm]
\{v(x),r(y)\}_1=(-p\p^2+p\p (H-h)-pv+r(\p -H+h))\dl(x-y),&\\[1mm]
\{v(x),w(y)\}_1=-q\dl(x-y),
{}~~~~~~~\{v(x),p(y)\}_1=(-p\p +r)\dl(x-y),&\\[1mm]
\{v(x),H(y)\}_1=-\frac{1}{2}(\p +H-h)\p \dl(x-y),~~~~~
\{v(x),h(y)\}_1=\frac{1}{2}(\p +H-h)\p \dl(x-y),&\\[1mm]

\{q(x),q(y)\}_1=-w\p w\dl(x-y),&\\[1mm]
\{q(x),r(y)\}_1=(\p^3-\p^2 H-H\p^2 +h_{xx}+2h_x\p +\p v+u\p
-H(u+v)-h(v-u)+&\\[1mm]
{}~~~~~~~~~~~~~~~~~~~~~~~~+(H+h)\p (H-h))\dl(x-y),
{}~~~~~~~\{q(x),w(y)\}_1=0,&\\[1mm]
\{q(x),p(y)\}_1=((\p -H-h)\p -v+u)\dl(x-y),~
{}~~~~\{q(x),H(y)\}_1=(\frac{1}{2}w\p -q) \dl(x-y),&\\[1mm]
\{q(x),h(y)\}_1=-\frac{1}{2}w\p \dl(x-y),~~~~~~~

\{r(x),r(y)\}_1=-p\p p\dl(x-y),&\\[1mm]

\{r(x),w(y)\}_1=((\p +H-h)\p +v-u)\dl(x-y),~~~~~
\{r(x),p(y)\}_1=0,&\\[1mm]

\{r(x),H(y)\}_1=(-\frac{1}{2}p\p +r) \dl(x-y),~~~~~~~
\{r(x),h(y)\}_1=-\frac{1}{2}p\p \dl(x-y),&\\[1mm]

\{w(x),w(y)\}_1=0,~~~~~~~~
\{w(x),p(y)\}_1=2(-\p+H)\dl(x-y),&\\[1mm]
\{w(x),H(y)\}_1=-w \dl(x-y),~~~~~~~
\{w(x),h(y)\}_1=0,~~~~~~~

\{p(x),p(y)\}_1=0,&\\[1mm]
\{p(x),H(y)\}_1=p \dl(x-y),~~~~~~~
\{p(x),h(y)\}_1=0,&\\[1mm]
\{H(x),H(y)\}_1=-\p \dl(x-y),
{}~~~~~~~\{H(x),h(y)\}_1=0,~~~~~~~~
\{h(x),h(y)\}_1=-\p \dl(x-y),&\\
\end{array}
\ee
and Hamiltonians may be calculated from
\be
{\cal H}_n=\frac{2}{n} \int{tr~res(L^{{\frac{n}{2}}})}dx,
{}~~~\forall n \geq 1
\ee
where {\em tr.} and {\em res} mean taking matrix trace and the
coefficient of the term $\p^{-1}$ respectively,
\be
{\bf L}=\p^2-\left[\ba{cc}H+h&w\\ p&-H+h\ea \right]\p+
\left[\ba{cc}u&q\\ r&v\ea \right].
\ee
\pagebreak

{\em Remark}:

The hierarchy may be rewritten by means of Hamiltonian
operator terminology: ${\bf u}_{t_{n}}={\bf B}_0 \dl({\cal H}_n)
 ={\bf B}_1 \dl({\cal H}_{n+1})$ with ${\bf B}_0 $ and ${\bf B}_1 $
 are $8\times 8$ matrix differential operators, which may read
off from the formulae(3.3-4).

\bigskip

Noticing another form of the spectral problem(3.1), which is
nothing but the matrix Schordinger problem, we may use the
standard theory\cite{feher1}\cite{Bil} to derive the
corresponding Poisson structures. It is straightforward to
check that the brackets(3.3) and (3.4) are the exactly the
 Gelfand-Dikii's first and second brackets.

The classical \w42 algebra comes into play with the following observation:
 If we do the reduction $h=0$ for the Poisson algebra(3.4),
we obtain
\be
\begin{array}{ll}
\{u(x),u(y)\}=({3\over4}\p^3-{3\over4}H\p H+u\p+\p u+
{3\over4}H_{xx}+{3\over2}H_x\p)\dl(x-y),&\\[1mm]
\{u(x),v(y)\}=(-w\p p-wr+qp-{1\over4}(\p^3-\p^2H-H\p^2
+H\p H))\dl(x-y),&\\[1mm]

\{u(x),q(y)\}=(-w\p^2-w\p H-wu+q\p +qH-{1\over4}(\p -
H)\p w))\dl(x-y),&\\[1mm]

\{u(x),r(y)\}=((\p -H)(\p p +r)+up-{1\over4}(\p -
H)\p p))\dl(x-y),&\\[1mm]

\{u(x),w(y)\}=(-w\p +q)\dl(x-y),~~~~~~~
\{u(x),p(y)\}=-r\dl(x-y),&\\[1mm]

\{u(x),H(y)\}=\frac{1}{2}(\p -H)\p \dl(x-y),&\\[1mm]

\{v(x),v(y)\}=({3\over4}\p^3-{3\over4}H\p H+v\p+\p v-
{3\over4}H_{xx}-{3\over2}H_x\p)\dl(x-y),&\\[1mm]

\{v(x),q(y)\}=({3\over4}\p^2 w +{3\over4}H\p w+\p q+
Hq+wv)\dl(x-y),&\\[1mm]

\{v(x),r(y)\}=-(p\p^2+{1\over4}\p^2 p-p\p H+pv+
{1\over4}H\p p-r\p +rH)\dl(x-y),&\\[1mm]

\{v(x),w(y)\}=-q\dl(x-y),
{}~~~~~~~\{v(x),p(y)\}=(-p\p +r)\dl(x-y),&\\[1mm]

\{v(x),H(y)\}=-\frac{1}{2}(\p +H)\p \dl(x-y),~~~~~~~

\{q(x),q(y)\}=-{3\over4}w\p w\dl(x-y),&\\[1mm]

\{q(x),r(y)\}=(\p^3-\p^2 H-H\p^2 +\p v+u\p -H(u+v)+
H\p H+{1\over4}w\p p)\dl(x-y),&\\[1mm]

\{q(x),w(y)\}=0,~~~~~~~\{q(x),p(y)\}=(\p -H)\p -
v+u)\dl(x-y),&\\[1mm]

\{q(x),H(y)\}=(\frac{1}{2}w\p -q) \dl(x-y),~~~~~~~

\{r(x),r(y)\}=-{3\over4}p\p p\dl(x-y),&\\[1mm]

\{r(x),w(y)\}=((\p -H)\p +v-u)\dl(x-y),~~~~~~~
\{r(x),p(y)\}=0,&\\[1mm]

\{r(x),H(y)\}=-\frac{1}{2}(p\p -r) \dl(x-y),~~~~~~~

\{w(x),w(y)\}=0,&\\[1mm]

\{w(x),p(y)\}=2(-\p+H)\dl(x-y),~~~~~~~
\{w(x),H(y)\}=-w \dl(x-y).
&\\[1mm]

\{p(x),p(y)\}=0,~~~~~~~
\{p(x),H(y)\}=p \dl(x-y),~~~~~~~
\{H(x),H(y)\}=-\p \dl(x-y),&\\
\end{array}
\ee
This algebra is nothing but the classical \w42 algebra(2.1)
with the fields redefinition
\be
T=u+v-H^2-pw,~~~~~~ v=v,~~~~~p=p,~~~~~w=w,~~~~~H=H,~~~~~q=q,~~~~~r=r.
\ee
Thus, we rediscover the classical \w42 algebra from the
viewpoint of integrable systems. Because of this equivelence, we call
the Poisson algebra(3.7) \w42 also. The explicit form of
integrable hierarchy associated with it can be read off
from the hierarchy(3.2). Here, we just give the first non
trivial flow
\be
\begin{tabular}{l}
$u_t={1\over2}(-HH_x+H_{xx}+2u_x-wp_x-wr+qp)$,\\[1mm]
$v_t={1\over2}(-HH_x+H_{xx}+2v_x-w_xp+wr+qp)$,\\[1mm]
$q_t={1\over2}(-Hw_x+w_{xx}+2q_x+wH_x-2Hq+wu-wv)$,\\[1mm]

$r_t={1\over2}(Hp_x+p_{xx}+2r_x-pH_x-pu+pv+2rH)$,\\[1mm]

$w_t=p_t=H_t=0$.\\
\end{tabular}
\ee
{\em Remark:}

We note that in the system(3.9) the time evolution of the
fields (w,p,H) is trivial. This means that the dynamical
system may be reduced to the submanifold of (u,v,q,r).
In fact, this is a general phenomenon: the whole
hierarchy(3.2) is reducible to the submanifold (u,v,q,r).

\section{THE FREE FIELD REALIZATION OF THE \w42 ALGEBRA}
\setcounter{equation}{0}
For a given W algebra, it is important yet interesting to
 construct free field realization. Next we construct such
realization for our \w42 algebra(3.7). To this end, we start
with the derivation of Miura maps for the related hierarchy.
\par
Let us do the following factorization
\be
{\bf L}=(\p -M)(\p -N)
\ee
where ${\bf L}$ is given by (3.6),
$M=\left[\ba{cc} g_1&k\\ l&g_2\ea\right]$,
$N=\left[\ba{cc} m_1&n\\ s&m_2\ea\right]$.
Then, the transformation between field variables,
which is a Miura map, reads
\be
\begin{tabular}{ll}
$u=g_1m_1+ks-m_{1x},$&$v=ln+g_2m_2-m_{2x},$\\[1mm]

$q=g_1n+km_2-n_x,$&$r=lm_1+g_2s-s_x,~~ ~~~~~~w=k+n,$\\[1mm]

 $p=l+s,~~~~~~~~H={1\over2}(g_1+m_1-g_2-m_2)$,& $
h={1\over2}(g_1+g_2+m_1+m_2),$
\end{tabular}
\ee
and the spectral problem for the modified hierarchy is
\be
\Psi_x=\left[\ba{cccc}m_1&n&\la&0\\ s&m_2&0&\la\\
\la&0&g_1&k\\ 0&\la&l&g_2\ea\right]\Psi.
\ee
As for the modified Poisson bracket, one may either
calculate the bracket directly following\cite{feher1}\cite{Bil}
or use (4.3) to calculate zero curvature equation. The resulted
bracket is defined by the Hamiltonian operator
\be
\hat{{\bf B}}_0=\left[\ba{cccccccc}-\p&0&n&-s&0&0&0&0\\
0&-\p&-n&s&0&0&0&0\\ -n&n&0&-\p +m_1-m_2&0&0&0&0\\
s&-s&-\p -m_1+m_2&0&0&0&0&0\\
0&0&0&0&-\p&0&k&-l\\ 0&0&0&0&0&-\p&-k&l\\
0&0&0&0&-k&k&0&-\p +g_1-g_2\\ 0&0&0&0&l&-l&-\p -
g_1+g_2&0 \ea\right],
\ee
it can be directly verified that the Miura map(4.2) is
 a Hamiltonian or Poisson map. That is, it maps the modified
 Poisson bracket, defined by(4.4), to the Poisson bracket(3.4).
Up to now, all these are known(see\cite{feher1}\cite{Bil}).
However, we note that unlike the scalar case, the Miura map(4.2)
does not supply us a free field realization for the Poisson
algebra(3.4) although it does simplify this algebra.
To obtain such a realization, we need to introduce
further coordinates transformations. Since the block
structure of the $\hat{{\bf B}}_0$, we only need to work on
the subspace$(m_1,m_2,n,s)$ with
\be
{\bf B}_{11}=\left[\ba{cccc}-\p&0&n&-s\\ 0&-\p&-n&s\\
-n&n&0&-\p +m_1-m_2\\ s&-s&-\p -m_1+m_2&0\ea\right],
\ee
we observe that the following transformation
\be
\bar{m}_1=m_1+m_2, ~~~~~~~\bar{m}_2=m_1-m_2, ~~~~~~~\bar{n}=n,
{}~~~~~~~\bar{s}=s,
\ee
maps (4.5) to
\be
\hat{{\bf B}}_{11}=\left[\ba{cccc}-2\p&0&0&0\\
0&-2\p&2\bar{n}&-2\bar{s}\\ 0&-2\bar{n}&0&-\p +\bar{m}_2\\
 0&2\bar{s}&-\p -\bar{m}_2&0\ea\right].
\ee
At this stage, we may use Wakimoto construction\cite{wak}
to simplify the structure(4.7) further. Thus, we have
\be
\bar{m}_1=\xi,~~~~~~~
\bar{m_2}=\sqrt{2}\alf +2\bi \gamma,~~~~~~~
\bar{n}=-\bi \gamma^2+\gamma_x-\sqrt{2} \gamma\alf,~~~~~~~
\bar{s}=\bi,
\ee
and the operator in this coordinate $(\xi,\alf,\gamma,\bi)$ is
\be
{\bf D}_{11}=\left[\ba{cccc}-2\p&0&0&0\\ 0&-\p&0&0\\ 0&0&0&-1\\
 0&0&1&0\ea\right].
\ee
Since the block structure of the operator(4.4), we only need to
do the exact same transformations(4.6) and (4.9) for the other
 block. That is,
\be
\bar{g}_1=g_1+g_2,~~~~~\bar{g}_2=g_1-g_2, ~~~~~\bar{k}=k,
{}~~~~~ \bar{l}=l
\ee
\be
\bar{g}_1=\zeta, ~~~~~\bar{g}_2=\sqrt{2}\mu+2\eta\nu,~~~~~
\bar{k}=-\nu\eta^2+\eta_x-\sqrt{2}\mu\eta, ~~~~~\bar{s}=\nu.
\ee
Then, the final bracket is defined by the operator in the
coordinate $(\xi,\alf,\gamma,\bi,\zeta,\mu,\eta,\nu)$
\be
{\bf D}=\left[\ba{cccccccc} -2\p&0&0&0&0&0&0&0\\
0&-\p &0&0&0&0&0&0\\ 0&0&0&-1&0&0&0&0\\
0&0&1&0&0&0&0&0\\ 0&0&0&0&-2\p&0&0&0\\ 0&0&0&0&0&-\p&0&0\\
0&0&0&0&0&0&0&-1\\
0&0&0&0&0&0&1&0\ea\right]
\ee
Thus, we reach the free field realization for the Poisson
algebra(3.4).

Now we turn to the classical \w42 algebra. Let us first do
the following recoordinating
\be
(m_1,m_2,n,s,g_1,g_2,k,l) \rightarrow (m_1,m_2,n,s,g_1,k,l,E)
\ee
where $E=m_1+m_2+g_1+g_2$.
Under this new coordinate, $\hat{{\bf B}}_0$ becomes
\be
\bar{\bf B}_0=\left[\ba{cccccccc}-\p&0&n&-s&0&0&0&-\p\\
0&-\p&-n&s&0&0&0&-\p\\ -n&n&0&-\p +m_1-m_2&0&0&0&0\\
s&-s&-\p -m_1+m_2&0&0&0&0&0\\ 0&0&0&0&-\p&k&-l&-\p\\
0&0&0&0&-k&0&-\p +g_1-g_2&0\\ 0&0&0&0&l&-\p -g_1+g_2&0&0\\
-\p&-\p&0&0&-\p&0&0&-4\p \ea\right]
\ee
where $g_2=E-m_1-m_2-g_1$.

Now we do the Dirac reduction $E=0$ for the structure
$\bar{\bf B}_0$, the resulted structure is

$\bar{\hat{\bf B}}_0=$
\be
\left[\ba{ccccccc}
-{3\over4}\p&{1\over4}\p&n&-s&{1\over4}\p&0&0\\
 {1\over4}\p&-{3\over4}\p&-n&s&{1\over4}\p&0&0\\
 -n&n&0&-\p +m_1-m_2&0&0&0\\ s&-s&-\p -m_1+m_2&0&0&0&0\\
 {1\over4}\p&{1\over4}\p&0&0&-{3\over4}\p&k&-l\\
0&0&0&0&-k&0&-\p+2g_1+m_1+m_2\\
0&0&0&0&l&-\p-2g_1-m_1-m_2&0 \ea\right]
\ee
We claim that the Poisson structure induced by
$\bar{\hat{\bf B}}_0$ is related to the algebra
\w42 (3.7) by the following transformation
\be
\begin{array}{ll}
u=g_1m_1+ks-m_{1x},&v=ln-m_2(m_1+m_2+g_1)-m_{2x},\\[1mm]

q=g_1n+km_2-n_x,&r=lm_1-s(g_1+m_1+m_2)-s_x\\[1mm]

w=k+n, ~~~~~~~~p=l+s,& H=g_1+m_1
\end{array}
\ee
This statement may be verified by tedious but straightforward
 calculation.
\par

As above, this realization does not qualify as a free
 field realization and we need simplify (4.15) further to
reach such position. Our observation is that the following
 coordinations transformation $(m_1,m_2,n,s,g_1,k,l)\rightarrow
(\hat{m}_1,\hat{m}_2,\hat{n},\hat{s},\hat{g}_1,\hat{k},\hat{l})$
\be
\begin{tabular}{ll}
$\hat{m}_1=m_1+m_2,$&$ \hat{m}_2=m_1-m_2,~~~~~~~\hat{n}=n,
{}~~~~~~~ \hat{s}=s,$\\[1mm]

$\hat{g}_1=2g_1+m_1+m_2$,&$\hat{k}=k,~~~~~~~\hat{l}=l,$
\end{tabular}
\ee
brings $\bar{\hat{{\bf B}}}_0$ into a block matrix operator:
\be
{\tilde B}_0=\left[\ba{ccccccc}-\p&0&0&0&0&0&0\\
0&-2\p&2\hat{n}&-2\hat{s}&0&0&0\\ 0&-2\hat{n}&0&-\p
+\hat{m}_2&0&0&0\\ 0&2\hat{s}&-\p -\hat{m}_2&0&0&0&0\\
0&0&0&0&-2\p&2\hat{k}&-2\hat{l}\\ 0&0&0&0&-2\hat{k}&0&-\p +
\hat{g}_1\\
0&0&0&0&2\hat{l}&-\p-\hat{g}_1&0\ea\right]
\ee
Interesting enough, once again, we may use the Wakimoto
 construction directly for the structure(4,18). It reads as
\be
\begin{array}{ll}
\hat{m}_1=\theta,~~~~~~\hat{m}_2=\sqrt{2}\theta_1+
2\theta_2\theta_3,~~~~~~\hat{n}=-\theta_{2}^2\theta_{3}+
\theta_{2x}-\sqrt{2}\theta_1\theta_2,
{}~~~~~~&\hat{s}=\theta_3,\\[1mm]

\hat{g}_1=\sqrt{2}\vartheta_1+2\vartheta_2\vartheta_3,
{}~~~~\hat{k}=-\vartheta_{2}^2\vartheta_{3}+\vartheta_{2x}-
\sqrt{2}\vartheta_1\vartheta_2,&
\hat{l}=\vartheta_3
\end{array}
\ee
and final operator in coordinate $(\theta,\theta_1,
\theta_2,\theta_3,\vartheta,\vartheta_1,\vartheta_2,
\vartheta_3)$ is
\be
{\bf D}_0=\left[\ba{ccccccc} -\p&0&0&0&0&0&0\\
0&-\p&0&0&0&0&0\\ 0&0&0&-1&0&0&0\\ 0&0&1&0&0&0&0\\
0&0&0&0&-\p&0&0\\ 0&0&0&0&0&0&-1\\ 0&0&0&0&0&1&0\ea\right]
\ee

Then, the composition of (4.16-17) with(4.19) supplys us
free field realization for the \w42 algebra(3.7).\par

{\em Remarks:}

(1). This construction provides us, as a by-product, a new
proof of the Hamiltonian nature of the structure(3.4);

(2).The modified hierarchies for each coordinates are
easily calculated;

(3).With the free field realizations, we may construct
quantized algebras for the Poisson algebra(2.1), (3.4) and(3.7).

\section{TODA TYPE THEORIES CONNECTED WITH $W_4^{(2)}$}
\setcounter{equation}{0}

In this section we shall construct the Toda type theory connected
with the $W_4^{(2)}$ algebra. Exactly speaking, we shall construct a
Toda theory which corresponds to two copies of the $W_4^{(2)}$ algebra:
one copy is holomorphic, the other is anti-holomorphic.
The construction is based on the following observations. Recall
that the $W$-basis of the holomorphic
copy of $W_4^{(2)}$ used in \cite{Bak} is
arranged in the following Drinfeld-Sokolov gauge,

\begin{eqnarray*}
Q=\left(
\begin{array}{cccc}
$$0$$ & $$0$$ & $$-1$$ & $$0$$ \cr
$$0$$ & $$0$$ & $$0$$  & $$-1$$ \cr
$$T_1$$ & $$G^{(+)}$$ & $$U$$ & $$Z$$ \cr
$$Y$$ & $$T_2$$ & $$G^{(-)}$$ & $$-U$$
\end{array}
\right).
\end{eqnarray*}

\noindent Similarly we can have a $W$-basis of the
anti-holomorphic copy of $W_4^{(2)}$ which can also be arranged
into the Drinfeld-Sokolov gauge

\begin{eqnarray*}
\bar{Q}=\left(
\begin{array}{cccc}
$$0$$ & $$0$$ & $$\bar{T}_1$$ & $$\bar{Y}$$  \cr
$$0$$ & $$0$$ & $$\bar{G}^{(+)}$$ & $$\bar{T}_2$$  \cr
$$-1$$ & $$0$$ & $$\bar{U}$$ & $$\bar{G}^{(-)}$$ \cr
$$0$$ & $$-1$$ & $$\bar{Z}$$ & $$-\bar{U}$$
\end{array}
\right).
\end{eqnarray*}

Let $g$ be the solution of the following linear systems,

\begin{displaymath}
\partial_+ g + Q g = 0,\hskip 0.5truecm
\partial_- g + g \bar{Q} = 0.
\end{displaymath}

\noindent We can easily see that the matrix $g$ can be realized
by the matrix elements

\begin{displaymath}
g_{a}^{b} = \sum_i f_a^i \bar{f}_i^b,
\end{displaymath}

\noindent where $f_{i}^{j}$ and $\bar{f}_i^j$ satisfy

\begin{eqnarray*}
\partial_x f_{a}^{j} = - f_{a+2}^{j},\hskip 0.5truecm
\partial_x \bar{f}_{j}^{b} = - \bar{f}_{j}^{b+2},\hskip 0.5truecm
a=1,\;2.
\end{eqnarray*}

\noindent and $\bar{f}_i^b$ have the similar property. Define
the main diagonal subdeterminants $\Delta_a$ of the matrix $g$,
{\it i.e.}

\begin{eqnarray*}
\Delta_a=\left|
\begin{array}{ccc}
g_1^1 & ... & g_1^a \cr
\vdots &    & \vdots \cr
g_a^1 & ... & g_a^a
\end{array}
\right|,
\end{eqnarray*}

\noindent and, in particular, $\Delta_0 \equiv 1$,
we can prove, by tedious but direct calculations, that
the matrix $T$ with the elements (here $\Delta_a(i,\;j)$ denotes the
algebraic co-minor of $\Delta_a$ with respect to $g_i^j$)

\begin{eqnarray*}
T_a^b \equiv \sqrt{\frac{\Delta_{a-1}}{\Delta_a}} \sum_{l=1}^a
\frac{\Delta_a (l,\;a)}{\Delta_{a-1}} f_a^b
\end{eqnarray*}

\noindent satisfy the following equations,

\begin{eqnarray}
\partial_\pm T = \pm \left( \frac{1}{2} \partial_\pm \Phi +
{\rm exp}( \mp \frac{1}{2}{\rm ad} \Phi) (\Psi_\pm + \mu_\pm ) \right) T,
\label{Lax}
\end{eqnarray}

\noindent where we used the following abbreviations of notations,

\begin{eqnarray}
& &\Phi = \sum_{i=1}^{3} \phi^i H_i,\hskip 0.5truecm
\phi^a = {\rm ln} \Delta_a,\nonumber\\
& &\Psi_+ = \sum_{j=1}^{3} \sum_{i=1}^{4} {\rm sign}(i-j)
\psi_i^+ A_{ij} E_j,\hskip 0.5truecm
\psi_a^+ = \frac{\Delta_{a+1}(a,\;a+1)}{\Delta_a},\nonumber\\
& &\Psi_- = \sum_{j=1}^{3} \sum_{i=1}^{4} {\rm sign}(i-j)
\psi_i^- A_{ij} F_j,\hskip 0.5truecm
\psi_a^- = \frac{\Delta_{a+1}(a+1,\;a)}{\Delta_a}, \label{Sol}
\end{eqnarray}

\noindent $H_i$, $E_i$ and $F_i$ are the standard Chevalley
generators of the Lie algebra $A_3$ written in the defining
representation, $A$ is the following matrix

\begin{eqnarray*}
A= \left(
\begin{array}{cccc}
2 & $$-1$$ & 0 & 0 \cr
$$-1$$ & 2 & $$-1$$ & 0 \cr
0 & $$-1$$ & 2 & 0 \cr
0 & 0 & $$-1$$ & 0
\end{array}
\right),
\end{eqnarray*}

\noindent and $\mu_\pm$ are defined as

\begin{eqnarray*}
& &\mu_+ = \frac{1}{2} \sum_{i,\;j=1}^{3} \left[ E_i,\;E_j \right] ,\\
& &\mu_- = - \frac{1}{2} \sum_{i,\;j=1}^{3} \left[ F_i,\;F_j \right] .
\end{eqnarray*}

Equation (\ref{Lax}) can be viewed as the Lax pair of $W_4^{(2)}$ Toda
theory, with the explicit solution of the Toda fields given by
equation (\ref{Sol}). The Toda field equation can be easily obtained
from the compatibility condition of the Lax pair (\ref{Lax}).
The result reads

\begin{eqnarray*}
& &\partial_+ \partial_- \Phi + \left[ {\rm e}^{{\rm ad} \Phi} (\Psi_-),
\; \Psi_+ \right] + \left[ {\rm e}^{{\rm ad} \Phi} (\mu_-),\; \mu_+ \right]
= 0,\\
& &\partial_- \Psi_+  - \left[ \mu_+,\; {\rm e}^{{\rm ad} \Phi}
(\Psi_-) \right] = 0,\\
& &\partial_+ \Psi_-  - \left[ {\rm e}^{{\rm ad} \Phi} (\Psi_+),\;
\mu_- \right] = 0.
\end{eqnarray*}

\noindent In terms of the component fields, the above equations read
($K$ is the Cartan matrix of $A_3$)

\begin{eqnarray*}
& &\partial_+ \partial_- \phi^j - \sum_{i,\;k=1}^4 {\rm sign}(i-j)
{\rm sign}(k-j) \psi_i^+ A_{ij} \psi_k^- A_{kj} \omega^j
+ \sum_{l=1\;l \neq j}^3 \omega^l \omega^j K_{ij} =0,\\
& &\partial_-\psi_j^+ - \sum_{k=1}^{4} {\rm sign}(k-j)
\psi_k^- A_{kj} \omega^j=0,\\
& &\partial_+ \psi_j^- - \sum_{k=1}^{4} {\rm sign}(k-j)
\psi_k^+ A_{kj} \omega^j=0,\\
& &\omega^j \equiv {\rm exp}(- \sum_{i=1}^{3} \phi^i K_{ij} ),
\hskip 0.5truecm (j=1,\;2,\;3)\\
& &\partial_- \psi_4^+ = \partial_+ \psi_4^- =0.
\end{eqnarray*}

\noindent {\it Remarks.}

(1). The above construction of Toda type theory is essentially an extension
of the technique of $W$-surfaces, which was first developed by Gervais and
Matsuo \cite{GervaisMatsuo} in the standared $W_N$ cases. Thus the
construction given here not only present the $W_4^{(2)}$ Toda equation
but also the $W_4^{(2)}$ surface in the sence of \cite{GervaisMatsuo}.

(2). Toda type equations associated with general $W_N^{(2)}$ algebras are
already studied by one of the authors (LC) and collaborators in several
papers \cite{ch}. However those equations restricted to the case of
$N=4$ lack the fields $\psi_4^\pm$, thus does dot really corresponds to
$W_4^{(2)}$ algebra. The present equations overcome this shortcoming.

(3). The functions $f_a^i$ and $\bar{f}_i^a$ can be shown to satisfy two
commuting families of classical exchange algebra for $a=1,\;2$. For example,
the holomorphic family of exchange algebra reads

\begin{eqnarray}
& &\left\{ f_a^i(x),\;f_b^j(y) \right\} = - \frac{1}{8} f_a^i (x) f_b^j (y)
{\rm sign}(x-y) + f_a^j (x) f_b^i (y) \left[ \theta (i-j) \theta(x-y) -
\theta(j-i) \theta(y-x) \right], \nonumber\\
& &a,\;b=1,2, \label{EX}
\end{eqnarray}

\noindent where

\begin{equation}
\theta (a-b) \equiv \left\{
\begin{array}{cc}
\frac{1}{2} & $$(a-b=0)$$ \cr
0 & $$(a-b <0)$$ \cr
1 & $$(a-b >0)$$
\end{array}
\right.
, \hskip 0.5truecm
{\rm sign}(a-b) = \theta(a-b)-\theta(b-a).
\end{equation}

\noindent Such exchange algebras can be used to reconstruct $W_4^{(2)}$
algebra since one can always write the $W$-basis of $W_4^{(2)}$ algebra
in terms of appropriate determinants consisted of the above functions.
This construction of $W$ algebras can also be extended to  any classical
$W_N^{(l)}$ algebras \cite{cw}. Since the classical exchange algebra
is the origin of quantum group, it may also be possible to
relate quantum $W$ algebras
and quantum groups in terms of a quantized version of such constructions.

(4). The canonical Poisson structure for the $W_4^{(2)}$ Toda fields can also
be obtained from the exchange relation (\ref{EX})
and the explicit solution (\ref{Sol}) of the field equations.

\section{CONCLUSIONS}

In this paper we constructed both the integrable evolution equations
and the corresponding Toda theory associated to the $W_4^{(2)}$ algebra.
Miura maps are presented in connection with the $W_4^{(2)}$ evolution
equations, which in turn give a free field realization of \w42 algebra.

Though the problems considered here is only a specific case of the
$W$-algebra--evolution equation--Toda system connections, the
constructions presented here again assures the widely adopted conjecture
that given a $W$-algebra there must exist an associated system of evolution
equations and a corresponding Toda theory.

Besides what have been considered in the main text of this paper, we would
like to mention that there are still some unsolved problems such as the
connection between the variables appeared in the evolution equations and the
Toda fields. As the \w42 algebra is much more complicated than the standard
$W_N$ series, one should feel reasonable that such connections are not
so straighforward as in the standard case.

\vspace{10pt}
{\bf{ACKNOWLEDGEMENT}}\par
One of us(QPL) would like to thank Drs. C.S. Xiong and K. Wu for the helpful
discussions. This work is supported by Natural National Science
 Foundation of China.\par
\vspace{.1in}
\bigskip

\smallskip
\par

\par

\small
\begin{thebibliography}{}

\bibitem{feher2} J. Balog, L. Feh\'er, L. O'Raifeartaigh, P. Forgacs and
	       A. Wipf, Ann. Phys. {\bf 203}(1990)76;

	      L. Feh\'er, L. O'Raifeartaigh, P. Ruelle, I. Tsutsui and
	       A. Wipf,  Phys. Rep. {\bf 222}(1992)1;

	      P. Bouwknegt, and K. Schoutens, Phys. Rep.{\bf 223}(1993)183.


\bibitem{Dic} L.A. Dickey, {\em Soliton equations and
	      Hamiltonian systems}, World Scientific, Singapore(1991);

{}~~ M.J. Ablowitz and H. Segur, {\em Solitons and inverse
	      scattering transforms}, SIAM, Philadalphia(1981).

\bibitem{zam}  A. B. Zamolodchikov, Theor. Math. Phys.{\bf65}(1985)1205;

	    ~~ A. B. Zamolodchikov and V. A. Fateev, Nucl. Phys.
	   {\bf B280[FS18]}(1987)644;

	  ~~ V. A. Fateev and S. L. Lykyanov, Int. J. Mod. Phys.
	   {\bf A3}(1988)507.


\bibitem{poly} A. M. Polyakov, Int. J. Mod. Phys. {\bf A5}(1990)833;

	   ~~ M. Bershadsky, Commun. Math. Phys. {\bf 139}(1991)71.

\bibitem{Bais} F. A. Bais, T. Tjin and P. van Driel, Nucl. Phys.
	    {\bf B357}(1991)632.

\bibitem{Bil} A. Bilal, Lett. Math. Phys. {\bf 32} (1994)103;

{}~~ A. Bilal, Nonlocal Matrix Generalizations of W - algebras, hep-th/9403197.

\bibitem{Bak} I. Bakas and D.A. Depireux,  in {\em Proc. of  the XXth Int.
	Conf. on Diffenertial Geometric Methods in


    ~~ Theoretical Physics}, New York, June  1991.

\bibitem{de} D.A. Depireux and P. Mathieu, Inter.
		 J. Mod. Phys. {\bf A7} (1992)6053.

\bibitem{qpl} Q.P. Liu, Phys. Lett. {\bf A187} (1994)373.  \par
 \bibitem{xcs} Q.P. Liu and C.S. Xiong, Phys. Lett. {\bf B327} 257(1994).\par
\bibitem{bor} M. Borona, Q.P. Liu and C.S. Xiong,
	       Bonn-Th-9417, SISSA-ISAS-118/94, AS-ITP-94-43,

	     ~~ hep-th/9408035.

\bibitem{pm} P. Mathieu, Phys. Lett. {\bf B208} (1988) 101.

\bibitem{bak1} I. Bakas, Commun. Math. Phys. {\bf 123} (1989) 627.

\bibitem{ge} J.-L. Gervais, Phys. Lett. {\bf B160} (1985) 277.

\bibitem{feher1} L. Feh\'er, J. Harnad and I. Marshall,
	    Commun. Math. Phys. {\bf 154} (1993) 181.

\bibitem{guil} E. Olmedilla, L. Martinez Alonso
	    and F. Guil, Nuovo Cim. {\bf 61B} (1981) 49.

\bibitem{wak} M. Wakimoto, Commun. Math. Phys. {\bf 104}  (1986) 604.

\bibitem{ch} B.-Y. Hou, L. Chao, Int. J. Mod. Phys. {\bf A7} (1993) 7105,

	~~L. Chao, Commun. Theore. Phys. {\bf 20} (1993) 221,

	~~L. Chao, B.-Y. Hou, Ann. Phys. (NY) {\bf 230} (1994) 1,

       ~~B.-Y. Hou, L. Chao, Int. J. Mod. Phys. {\bf A8} (1993) 1105.

\bibitem{cw} L. Chao, Y.-S. Wang, preprint NWU-IMP (in preparation)

\bibitem{GervaisMatsuo} J.-L. Gervais, Y. Matsuo,
	   Commun. Math. Phys. {\bf 152} (1993) 317.

\end{thebibliography}
\end{document}